\documentclass[sigconf, authorversion]{acmart}

\usepackage{url}
\usepackage{graphicx}
\usepackage{booktabs}
\usepackage{tabularx}
\usepackage{booktabs}
\usepackage{balance}

\begin{document}

\title{Invisible Pixels Are Dead, Long Live Invisible Pixels!}

\author{Jukka Ruohonen}
\affiliation{\institution{University of Turku, Finland}}
\email{juanruo@utu.fi}

\author{Ville Lepp\"anen}
\affiliation{\institution{University of Turku, Finland}}
\email{ville.leppanen@utu.fi}

\begin{abstract}
Privacy has deteriorated in the world wide web ever since the 1990s. The tracking of browsing habits by different third-parties has been at the center of this deterioration. Web cookies and so-called web beacons have been the classical ways to implement third-party tracking. Due to the introduction of more sophisticated technical tracking solutions and other fundamental transformations, the use of classical image-based web beacons might be expected to have lost their appeal. According to a sample of over thirty thousand images collected from popular websites, this paper shows that such an assumption is a fallacy: classical $1 \times 1$ images are still commonly used for third-party tracking in the contemporary world wide web. While it seems that ad-blockers are unable to fully block these classical image-based tracking beacons, the paper further demonstrates that even limited information can be used to accurately classify the third-party $1 \times 1$ images from other images. An average classification accuracy of $0.956$ is reached in the empirical experiment. With these results the paper contributes to the ongoing attempts to better understand the lack of privacy in the world wide web, and the means by which the situation might be eventually improved.
\end{abstract}

%

%
\setcopyright{none}
\acmConference[WPES'18]{2018 Workshop on Privacy in the Electronic Society}{October 15, 2018}{Toronto, ON, Canada}
\acmBooktitle{2018 Workshop on Privacy in the Electronic Society (WPES'18), October 15, 2018, Toronto, ON, Canada}
\acmDOI{10.1145/3267323.3268950}

%



\maketitle

\section{Introduction}

In the early days of the Internet the term ``web bug'' referred to simple techniques with which unobtrusive user tracking was implemented in the world wide web. Together with web cookies, which were first standardized in 1997, these web bugs were an important historical factor shaping the developments that have continued to deteriorate privacy in the world wide web for two decades~\text{\cite{Lerner16, West17}}. Typically these web bugs were one-pixel-by-one-pixel images embedded to a given website but hosted from a different third-party website. These image-based characteristics also introduced concepts such as ``invisible image'', ``tracking pixel'', ``invisible pixel'', and ``pixel tag'' into the rubric of popular Internet discourse. The tracking itself was technically simple. When a user visited a website with a web bug, the loading of the image then delivered tracking information to the associated third-party once the user's client retrieved the image. This simple mechanism for client-side tracking is essentially the same today as it was about two decades ago.

The web advertisement industry and associates managed to later brand the term web bug with the more positive ``web beacon'' term. The same parties were also busy developing more sophisticated tracking techniques throughout the decades. The examples include so-called browser fingerprinting, multi-device identification, tracking through local storage and caching, canvas-based tracking, and Flash cookies \cite{EstradaJimenez17, LeFallace17}. 
Given the traditionally extensive lobbying at the World Wide Web Consortium \cite{Gamalielsson17, McDonald18}, also many standardization initiatives have been proposed as alternatives to image-based web beacons. For instance, a \texttt{ping} attribute has been introduced for \texttt{<a>} tags in order to make tracking easier; when a user clicks a hyperlink, a hypertext transfer protocol (HTTP) POST request is sent to the destination specified in the attribute. Analogously, a whole application programming interface has recently been proposed for web beacons~\cite{W3C18a}. Furthermore, the functionality of most popular websites nowadays depend on numerous third-parties \cite{Butkiewicz11, Ruohonen18PST}. Images, multimedia content, fonts, JavaScript libraries, style sheets, and many other web resources are commonly hosted on different third-party domains and delivered via content delivery networks.

All these fundamental transformations would lead one to expect that traditional image-based web beacons would have long-lost their appeal for third-party tracking. Already because most current websites load tens of web resources from third-parties and JavaScript provides overwhelming functionalities, the rationale for image-based beacons seems somewhat senseless in the contemporary world wide web. Once upon a time, invisible images had a web development function for styling websites \cite{Shannon12}, but those days are long gone. However, the forthcoming results show that this reasoning is false. The following three contributions are thus made:
\begin{enumerate}
\itemsep0.2em 
\item{Excluding some rare exceptions~\cite{Englehardt18, Krammer08}, there exists a very limited amount of empirical research on classical $1 \times 1$ third-party tracking images; the paper presents the supposedly first measurement study concentrating solely on this topic.}
\item{In contrary to prior expectations, the paper shows that image-based web beacons are still quite frequently used in 2018.}
\item{The paper demonstrates that third-party $1 \times 1$ images can be classified with a high accuracy even with limited data.}
\end{enumerate}

The structure of the paper's remainder is simple: the empirical results are presented in Section~\ref{section: results} and briefly discussed in Section~\ref{section: discussion}.

\clearpage
\pagebreak
\section{Results}\label{section: results}

The forthcoming results are disseminated in three straightforward steps: a brief elaboration of the dataset is followed by a few descriptive statistics, after which the classification results are presented.

\subsection{Data}

The dataset is based on the Alexa's ranking of the top-500 most popular websites in the global Internet \cite{Alexa18a}. While often used in Internet measurement research~\text{\cite{AlQudah10, Buchanan17, Falahrastegar14, StarovNikiforakis18}}, the list is small for probing image-based beacons. For this reason, (a)~the sampling procedure was implemented by visiting all hyperlinks present in the primary web pages of the domains in the top-500 list, provided that these shared the same second-level domain names. These additional visits based on the \texttt{href} attributes of the \texttt{<a>} tags were then mapped back to the domain names in the Alexa's list. It is important to further note that (b) all queries were initiated with plain HTTP, but (c) all redirections were followed. Following existing research~\cite{Ruohonen17EISIC, Ruohonen18IFIPSEC}, (d) all queries were made with a custom JavaScript-capable WebKit/Qt-powered web browser, and (e)~a $30$ second timeout was used for each query in order to ensure that full contents were loaded. Finally, (f) the top-500 list was processed three times to rule out temporary network failures. The point about Java\-Script is particularly important because many web beacons either require JavaScript or these are only visible for queries made with JavaScript. For instance, a fairly typical way for trying to hide image-based beacons is to use a zero-width and zero-height \texttt{<frame>} to which numerous image-based beacons are embedded with \texttt{<img>} tags.

Images were collected with HTTP GET requests from the \texttt{<img>} tags present in the web pages visited. Only images with unique cryptographic hashes were qualified to the sample on per-website basis. Thus: if a website referenced the same image with multiple \texttt{<img>} tags, the corresponding image is counted only once for this particular website. In terms of parsing, a library \cite{magic} for identifying multipurpose Internet mail extension (MIME) types was used to deduce about the scalable vector graphics (SVG) format. Images with MIME types other than \texttt{image/svg} and \texttt{image/svg+xml} were then passed to another library \cite{pillow}, and qualified to the analysis if the library recognized the images. An image is then defined as ``invisible'' when both the width and height equal one pixel. For SVG images, invisibility is defined to occur when the \texttt{width} and \texttt{height} attributes in the \texttt{<svg>} tags are equal or less than one.

The \texttt{src} attribute in a \texttt{<img>} tag is essential for deducing whether an invisible $1 \times 1$ image is used for tracking. While the so-called same-origin policy~\cite{Ruohonen18IFIPSEC, Schwenk17} could be used also for this task, a~more relaxed definition is adopted: an image is defined as ``cross-domain'' when the second-level domain name extracted from the uniform resource locator (URL) in the \texttt{src} attribute differs from the second-level domain name of the web page visited. As redirections are followed, the comparison is done according to the visited (and not requested) pages. Although the definition is a simplification~\cite{Butkiewicz11}, it is commonly used~\cite{Eravuchira16, Ruohonen18PST} and adequate for the paper's purposes.

\subsection{Descriptive Statistics}

In total, about 98\% of the five hundred domains sampled were successfully queried. These queries resulted over $30$ thousand images collected from \texttt{<img>} tags alone. From these, only a few were $1 \times 1$ images. The absolute amounts are deceiving, however. From the $488$ domains successfully queried, as many as $149$, or about 31\%, included at least one invisible image. The numerical details shown in Table~\ref{tab: sample} indicate that not all of these were used for third-party tracking, however. About 27\% of the URLs in the \texttt{src} attributes of the \texttt{<img>} tags referencing $1 \times 1$ images do not satisfy the given cross-domain definition. As there is little reason to nowadays use invisible images for purposes other than tracking, it is highly probable that this subset is used for ``cross-subdomain'' tracking across complex web deployments. The logic of tracking remains the same either way, but in these cases ``the third party is also a first party''~\citep{Mayer12} because a same entity controls the subdomains. When tracking clients visiting \texttt{example.com}, beacons may be sent to a dedicated \texttt{tracker.example.com}, for instance. Another point to make from Table~\ref{tab: sample} is that the share of cross-domain images is roughly comparable to the share of cross-domain $1 \times 1$ images. In other words, images are commonly hosted on third-party domains in general.

\begin{table}[th!b]
\centering
\caption{Sample Characteristics}
\label{tab: sample}
\begin{tabular}{lr}
\toprule
Domains sampled successfully & $488$ \\
All images from \texttt{<img>} tags & $30572$ \\
\quad $\bullet$ From which cross-domain images & 17760 \\
\quad $\bullet$ From which $1 \times 1$ images & $324$ \\
\quad $\bullet$ From which $1 \times 1$ cross-domain images & $235$ \\
\bottomrule
\end{tabular}
\end{table}

\begin{figure}[th!b]
\centering
\includegraphics[width=\linewidth, height=3cm]{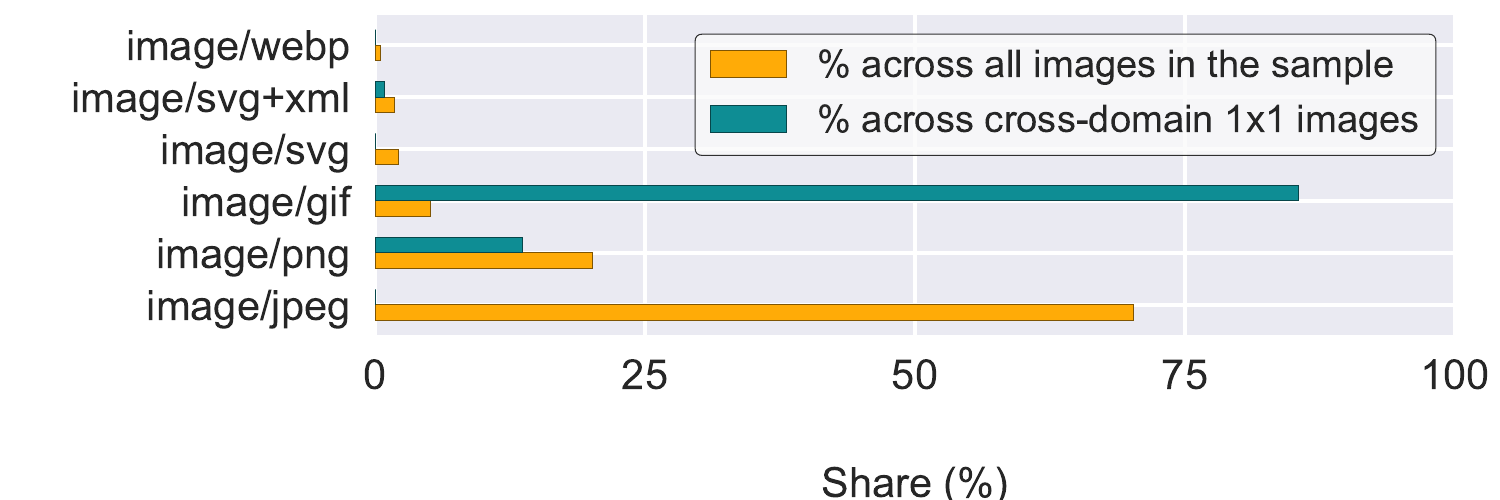}
\caption{MIME Types of the Images}
\label{fig: mime}
\end{figure}

\begin{figure}[th!b]
\centering
\includegraphics[width=\linewidth, height=4cm]{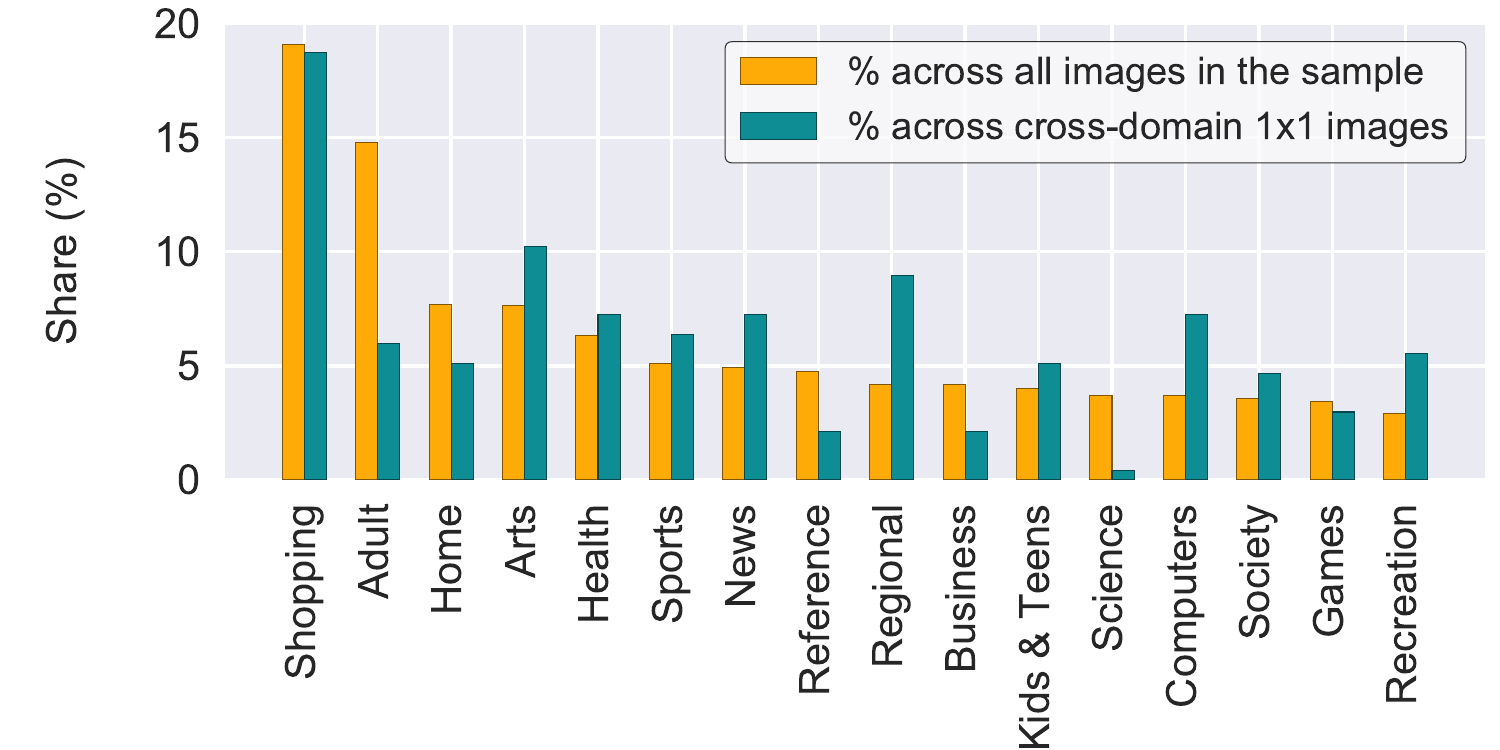}
\caption{Images Across Alexa's Top-500 Domain Categories}
\label{fig: categories}
\end{figure}

\begin{figure}[th!b]
\centering
\includegraphics[width=\linewidth, height=5.1cm]{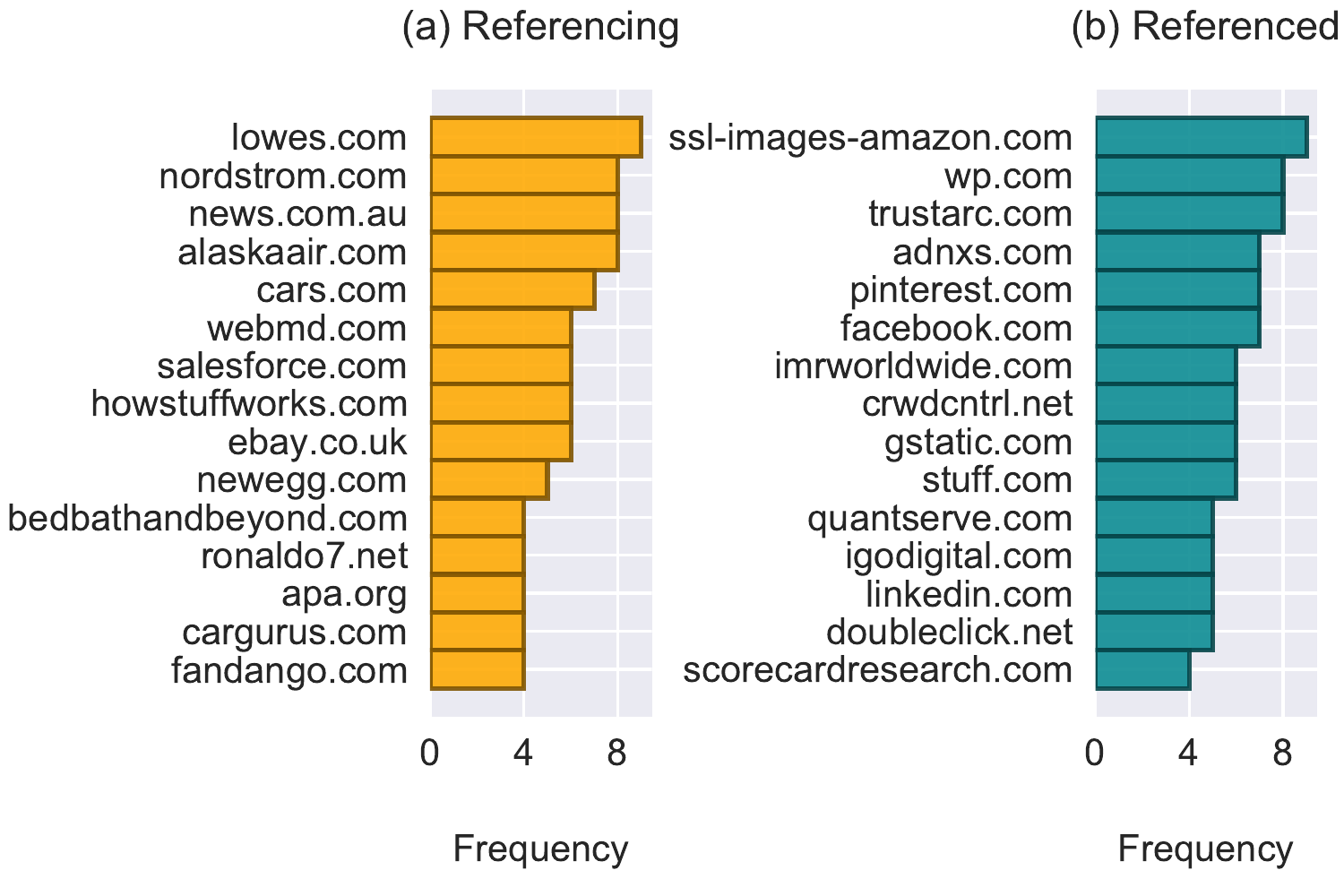}
\caption{The Top-Fifteen Referencing Domains and Referenced Second-Level Domains for $1 \times 1$ Cross-Domain Images}
\label{fig: domains}
\end{figure}

The MIME types summarized in Fig.~\ref{fig: mime} show no big surprises. The JPEG standard is the most common one for images in general, while most of the cross-domain $1 \times 1$ images are delivered in the graphics interchange format (GIF). That said, also the portable network graphics (PNG) format was used for a few of the invisible images. A similar breakdown according to the Alexa's genre categories for the top-500 websites is more interesting (see Fig.~\ref{fig: categories}). Although the media industry in general and newspaper websites in particular are often seen as particularly problematic for privacy due to the extensiveness of web advertisements~\cite{Englehardt16, Ruohonen17EISIC, StarovNikiforakis18, Traverso17}, the shopping genre attains the highest relative shares of both images and invisible images in the sample. Pornography websites expectedly contain many images, although the share of cross-domain $1 \times 1$ images is on a par with the sample average. The genres for arts and computers as well as popular regional websites are also noteworthy for their heavy use of image-based cross-domain beacons. The science genre attains the lowest relative share in the sample.

Before turning into the classification results, it is illuminating to take a peek at the domains using image-based beacons and the third-party domains from which these are loaded by clients. The left-hand side plot in Fig.~\ref{fig: domains} explains the large share of cross-domain $1 \times 1$ images in the shopping genre; many of the e-commerce sites are using multiple invisible images. Even though websites related to science are only infrequently using image-based beacons, it is worth pointing out the outlier pointing to the website of the American Psychological Association (APA). When turning to the right-hand side plot, it is evident that many of the conventional big players are using image-based beacons. The examples include Amazon, WordPress.com, Google, and LinkedIn, as well as many well-known web advertisement companies and their trackers. While Facebook has long been on the spotlight for its JavaScript-based beacons \cite{Williams18}, the results indicate that the company is using also traditional image-based tracking. All in all, these descriptive observations provide a good basis for defining features for classification.

\subsection{Classification Results}

The supervised learning experiment is implemented by classifying the $235$ cross-domain $1 \times 1$ tracking images against all other images collected. As the setup is highly unbalanced (see Table~\ref{tab: sample}), the experiment is carried out by random under-sampling from the majority class of normal, non-invisible images. While many alternatives are available~\cite{Tahir12}, this basic under-sampling works well enough in many applied problems. Thus, a $10$-fold cross-validation is carried out for $250$ random balanced samples. Arithmetic mean and standard deviation are used for reporting the results. A readily available decision-tree classifier \cite{scikitlearn} is used for the computation.

\begin{table}[th!b]
\centering
\caption{Features for Classification}
\label{tab: features}
\begin{tabularx}{\linewidth}{lX}
\toprule
Name & Description \\
\hline
QURL & True if a \texttt{query} field is present in a URL for an image. \\
\cmidrule{2-2}
QDOM & True if QURL is true \textit{and} any of the referencing domains (cf.~Fig.~\ref{fig: domains}) appear in a \texttt{query} field of an URL. \\
\cmidrule{2-2}
UNUM & Number of numbers $(0, \ldots, 9)$ appearing in an URL. \\
\cmidrule{2-2}
CORG & True if an image's URL is not only cross-domain but also cross-origin with respect to a sampled domain. \\
\cmidrule{2-2}
BLCK & True if an image's URL would be blocked by an ad-blocker \cite{EasyList18} according to an offline parser~\cite{adblockparser18}. \\
\cmidrule{2-2}
AALT & True if an \texttt{alt} attribute is present in a \texttt{<img>} tag. \\
\cmidrule{2-2}
ASTY & True if a \texttt{style} attribute is present in a \texttt{<img>} tag. \\
\cmidrule{2-2}
ETAG & True if an \texttt{Etag} field is set in a HTTP response. \\
\cmidrule{2-2}
COOK & True if a \texttt{Set-Cookie} is set in a HTTP response. \\
\cmidrule{2-2}
NOCH & True if \texttt{no-cache}, \texttt{no-store}, or \texttt{must-revalidate} is specified for a \texttt{Cache-Control} in a HTTP response. \\
\cmidrule{2-2}
MAGE & If present, the \texttt{max-age} value specified for a \texttt{Cache-Control} HTTP response field; $-1$ otherwise. \\
\cmidrule{2-2}
MIME & A dummy variable for five MIME types in Fig.~\ref{fig: mime} \\
\cmidrule{2-2}
DTOP & A dummy variable for the top-5 referenced domains, as listed in descending order in the plot (b) in Fig.~\ref{fig: domains}. \\
\bottomrule
\end{tabularx}
\end{table}

The features used for the classification experiment are enumerated in Table~\ref{tab: features}. These can be grouped analytically into four categories. The first category deals with the URLs extracted from the \texttt{src} attributes of the \texttt{<img>} tags used in the websites successfully sampled. These features are easy to justify based on existing research. For instance, the presence of a query field (QURL) is often associated with cross-origin JavaScript content prone to change temporally \cite{Ruohonen18IFIPSEC}. Another example would UNUM, which approximates the prevalence of identifiers embedded to URLs. Although reverse engineering is difficult, such identifiers are presumably used for tracking unique clients, users, or both \cite{Falahrastegar16, Mayer12}. Also the second group of features convey a clear rationale. For instance, there is little reason beyond obfuscation to specify an \texttt{alt} attribute for an invisible image. In contrast, a \texttt{style} attribute may be used to additionally specify that ``\texttt{width:~1px; height: 1px}'' or that ``\texttt{display:~none~!important}''. The third group contains five features that are all based on the HTTP header responses that were received upon retrieving the images based on GET requests. The rationale is again relatively clear-cut. As an example: to be efficient, image-based tracking beacons should disable client-side caching; hence, ETAG and NOCH should be false and true, respectively. The fourth and final group contains two features that both expand to sets of dummy variables. The dummy variables for the MIME types should provide some discriminate power due to the distribution shown in Fig.~\ref{fig: mime}. For instance, performance should clearly improve with the dummy variable that takes a value one for \texttt{image/gif}-based images and zero otherwise. The five dummy variables used for the DTOP feature are included as additional statistical controls.

Given these notes, the results are summarized in Table~\ref{tab: classification}. The first panel indicates that BLCK and DTOP alone hardly improve the classification performance. (Due to the under-sampling, a random classifier attains an accuracy rate of $0.5$.) This observation provides weak support for existing results regarding the inadequacy of most ad-blockers particularly when the context is expanded toward client-side tracking in general \cite{StarovNikiforakis18, Traverso17}. This point applies particularly to the so-called EasyList~\cite{EasyList18} used to define the BLCK feature \cite{Bhagavatula14, Englehardt16, Merzdovnik17, Wills16}. When all features from Table~\ref{tab: features} are included, however, the average accuracy is as high as $0.956$. Given the limited information used, this level of classification performance is exceptionally good. With more features and larger datasets to learn from, accuracy could be probably pushed even toward the $0.99$ range. 

\begin{table}[th!b]
\centering
\caption{Classification Results}
\label{tab: classification}
\begin{tabular}{lccccc}
\toprule
& \multicolumn{2}{c}{BLCK and DTOP only}
&& \multicolumn{2}{c}{All features} \\
\cmidrule{2-3}\cmidrule{5-6}
& Mean & Std.~dev. && Mean & Std.~dev. \\
\hline
Recall & $0.569$ & $0.046$ && $0.956$ & $0.030$ \\
Precision & $0.669$ & $0.101$ && $0.958$ & $0.029$ \\
Accuracy & $0.569$ & $0.046$ && $0.956$ & $0.030$ \\
\bottomrule
\end{tabular}
\end{table}

As the ratio of cross-domain $1 \times 1$ images to all images may not be stable when also less popular domains are sampled, further empirical experiments are required, however. Due to the general limitations of domain name popularity lists~\cite{Scheitle18a}, it should be remarked that merely scanning a larger list does not solve the root issue. The results reported may also contain some inaccuracies since the so-called public suffix list~\cite{SuffixList18a} was not used for domain name comparisons. As it is unclear what the suffixes mean in terms of the domain name system, the use of the list remains debatable, however. Although empirical observations indicate otherwise \cite{Englehardt16}, it may be also possible that dynamically generated web content delivered to a client vary according to the client's geographic location. A~more fundamental question is whether the cross-domain definition used makes sense because the actual delivery is often close to the client due to content delivery networks \cite{Ruohonen18PST}. It is a much bigger question whether and to which extent such ``third-party'' networks conduct client tracking and profiling. However, surveillance rather than privacy is arguably a better concept for approaching this question.

\section{Discussion}\label{section: discussion}

Classical cross-domain $1 \times 1$ image beacons are still frequently used in the world wide web. Even limited information and a basic machine learning approach can classify such images to a very high level of accuracy. While these observations are interesting and noteworthy on their own right, these allow to also contemplate a little about the state of the current web privacy research.

Much of the research in the domain---including this paper---has been preoccupied with demonstrations that different forms of third-party tracking are prevalent. Another common topic has been the (in)efficiency of ad-blockers to counter such tracking. Although these empirical demonstrations are important from a viewpoint of regulation and policy-making \cite{Degeling18}, less attention has been given for engineering innovative countermeasures. The lack of robust countermeasures may also explain the continuing use of invisible tracking images. Although there may be technical reasons to prefer these images for email tracking~\cite{Englehardt18}, arguably only the human imagination limits the amount of plausible alternatives for web tracking. Against this backdrop, the explanation for the use of invisible images and other ``legacy techniques'' may be simple: why deprecate something that already works? Another explanation may be that these techniques provide a ``backup solution'' for countering simple countermeasures such as per-website JavaScript restrictions.

Even though some skepticism has been expressed about machine learning approaches~\cite{Merzdovnik17}, the classification results presented are promising. These also support existing observations about high accuracy rates in the ad-blocking context \cite{Bhagavatula14, Krammer08, Lashkari17, OMeara18}. Although an early machine learning application was published already almost two decades ago~\cite{Kushmerick99}, practically all of the ad-blocking solutions in day-to-day use are based on blacklists, messy regular expressions, and manual maintenance. The drawbacks are thus clear. Obviously, classifying invisible images provides only a very limited viewpoint on third-party tracking, but, on the other hand, the same goes for the commonly examined third-party JavaScript content referenced with \texttt{<script>} tags. In general, the problem is that practically all web elements that allow referencing external content may be used for third-party tracking. Many of these elements allow to also reference further external resources; URLs can be embedded to style sheets, JavaScript to SVG images, and so forth. 

Thus, it seems futile to even try to win the race about new tracking techniques. Instead, a potential machine learning approach for privacy-conscious users might be much simpler: it might be possible to build a reasonable training set with existing ad-blocking lists and other common countermeasures, and then simply instrument all cross-domain HTTP requests made within a browser. For such users, some amount of false positives is acceptable, and per-website exceptions can be always added through a user interface.

A further gap in the current web privacy research relates to the limited understanding on the effectiveness of personalized web advertisements and third-party tracking in general. In this regard, the situation is somewhat paradoxical for practical privacy research: to engineer machine learning solutions for better privacy protection, a better understanding is required about the efficiency of current tracking solutions, which, in turn, requires (reverse) engineering privacy-violating prototypes. For instance, browser fingerprinting might be countered with random shuffling of user-agent strings, small jitter introduced to the font and screen sizes announced by a browser, and other types of randomization \cite{Laperdrix15, Nikiforakis15}. But before implementing such solutions for actual users, a better understanding is required about the effectiveness of current browser fingerprinting techniques used in the wild. While progress has been made~\cite{Vastel18}, many of the questions are still unclear particularly when extended to fingerprinting beyond browsers. Finally, it remains worth asking whether the current web privacy research is seeing the forest for the trees? Do new tracking techniques matter in practice when the authors of this paper are likely identifiable with a few web beacons, Internet protocol addresses, and search engine histories? 

\balance
\clearpage
\pagebreak
\bibliographystyle{abbrv}

\end{document}